\documentclass[11pt,titlepage]{article}

\usepackage{setspace} 
\usepackage[english]{babel}
\usepackage[intlimits]{amsmath}
\usepackage[dvips]{graphicx}
\usepackage{amsmath}
\usepackage[round,numbers,sort&compress]{natbib} 

\title{ATP hydrolysis stimulates large length fluctuations in single actin filaments}
\author{Evgeny B. Stukalin\\
        Department of Chemistry,\\
        Rice University, Houston, TX 
\and   Anatoly B. Kolomeisky\thanks{ Corresponding author.  Address: 
                                     Department of Chemistry,
                                     Rice University, Houston, TX 77005 USA,
                                     Tel.:~(713)348-5672, Fax:~(713)348-5155}\\
                                     Department of Chemistry and Department of Chemical and Biomolecular Engineering,\\
                                     Rice University, Houston, TX }

\date{}

\begin{document}

\maketitle

\abstract{Polymerization dynamics of single actin filaments is investigated theoretically using a  stochastic model that takes into account the hydrolysis of ATP-actin  subunits, the geometry of actin filament tips, the lateral interactions between the monomers as well as the processes at both ends of the polymer. Exact analytical expressions are obtained for a mean growth velocity and for  dispersion in length fluctuations. It is found that the ATP hydrolysis has a strong effect on dynamic properties of single actin filaments. At high concentrations of free actin monomers the mean size of unhydrolyzed  ATP-cap is very large, and the dynamics is governed by association/dissociation of ATP-actin subunits. However, at low concentrations the size of the cap becomes finite, and the dissociation of ADP-actin subunits makes a significant contribution to overall dynamics. Actin filament length fluctuations reach the maximum at the boundary between two dynamic regimes, and this boundary is always larger than the critical concentration. Random and vectorial mechanisms of hydrolysis are compared, and it is found that they predict qualitatively similar dynamic properties. The possibility of attachment and detachment of oligomers  is also discussed.  Our theoretical approach is successfully applied to analyze the latest experiments on the growth and length fluctuations of individual actin filaments.}     


\clearpage

\section*{Introduction}

Actin filaments are major component of cytoskeleton in eukaryotic cells, and they  play important roles in many biological processes, including the organization of cell structures, transport of organelles and vesicles, cell motility, reproduction and endocytosis \citep{howard_book, bray_book, pollard03}. Biological functions of actin filaments are mostly determined by the dynamic processes that take place during the growth or shrinking of these biopolymers. However, our understanding of mechanisms of assembly and disassembly of these filaments is still very limited.

In recent years the number of experimental investigations of the growth dynamics of rigid cytoskeleton filaments, such as actin filaments and microtubules, at a single-molecule level have increased significantly \citep{erickson92, desai97, dogterom97, shaw03, kerssemakers03, janson04, lehto03,fujiwara02,kuhn05}. Dynamic behavior of individual microtubules have been characterized by a variety of experimental techniques such as video and electron microscopy, fluorescence spectroscopy, and optical trap spectrometry \citep{erickson92, desai97, dogterom97, shaw03, kerssemakers03, janson04}, whereas the studies of the single actin filaments have just started \citep{lehto03,fujiwara02,kuhn05}. The assembly dynamics of individual actin filaments revealed a treadmilling phenomenon, i.e., the polymer molecule tends to grow at the barbed end and to depolymerize at the pointed end \citep{fujiwara02}. Similar picture has been observed earlier for microtubules \citep{hotani88, grego01}.  Although the conventional actin filaments do not exhibit the dynamic instability as observed in microtubules  \citep{hotani88, mitchison92}, it was shown recently that the DNA-segregating prokaryotic actin homolog  ParM displays two phases of polymer elongation and shortening  (\citep{garner04}). ATP hydrolysis is not required for actin assembly \citep{cruz00}, but  it is known to play an important role in the actin polymerization dynamics. Experimental observations suggest that the nucleotide bound to the actin filament acts as a timer to control the filament turnover during the cell motility \citep{pollard00}. Hydrolysis of ATP and release of inorganic phosphate are assumed to promote the dissociation of the filament branches and the disassembly of ADP-actin filaments \citep{blanchoin00}. 

Recent experimental studies of the single actin filament growth \citep{fujiwara02,kuhn05} have revealed unexpected properties of actin polymerization dynamics. A large discrepancy in the kinetic rate constants for actin assembly estimated by average length change in the initial polymerization phase and determined from the analysis of length fluctuations in the steady-state phase (by a factor of 40) has been observed. Several possible explanations of this intriguing observation has been proposed \citep{fujiwara02,kuhn05,stukalin05,vavylonis05}. First, the actin polymerization dynamics might involve the assembly and disassembly of large oligomeric actin subunits. However, this point of view contradicts the widely accepted picture of single-monomer polymerization kinetics \citep{howard_book, vavylonis05}. In addition, as we argue below, it would require the association/dissociation of actin oligomers with 30-40 monomers, but the annealing of such large segments has not been observed in experiments  or has been excluded from analysis \citep{fujiwara02, kuhn05}. Second, stochastic pauses due to filament-surface attachments could increase the apparent dispersion in length of single actin-filaments, although it seems that the effect is not significant \citep{kuhn05}. Third, the errors in the experimental measurements could contribute into the observation  of large apparent diffusion constants \citep{kuhn05}. Another possible reason for the discrepancy is the use of oversimplified theoretical model in the analysis that neglects the polymer structure and the lateral interaction in the actin filament. However, the detailed theoretical investigation of the growth of single actin filaments \citep{stukalin05} indicates that large length fluctuations still cannot be explained by correctly describing the structure of the filament's tip and the lateral interactions between the monomers.

The fact that hydrolysis of ATP bound to the actin monomer is important  stimulated a different model to describe the actin polymerization dynamics \citep{vavylonis05}. According to this approach,  the ATP-actin monomer in the filament can be irreversibly hydrolyzed and transformed  into the ADP-actin subunit. The polymer growth is  a process of adding single ATP-actin monomers and deleting hydrolyzed or unhydrolyzed subunits, and the actin filament consists of two parts - a hydrolyzed core in the middle and unhydrolyzed caps at the polymer ends.  Large length fluctuations are predicted near the critical concentration for the barbed end.  Although this model provides a reasonable description of the filament growth rates for different ATP-actin concentrations, the position of the peak in dispersion is below the critical concentration for the barbed end, while in the experiments  \citep{fujiwara02, kuhn05} large dispersion is observed at, or slightly above,  the critical concentration. In addition, the proposed analytical method (without approximations) \citep{vavylonis05} cannot calculate analytically cap sizes and dispersion.  

The goal of this work is to develop a theoretical model of polymerization of single actin filaments that incorporates the ATP  hydrolysis in the polymer, the structure of the filament tips, lateral interactions between the monomers and dynamics at both ends. Our theoretical method is  based on  the stochastic models developed for describing the growth dynamics of rigid multifilament biopolymers \citep{stukalin04, stukalin05}, and it  allows to calculate explicitly all dynamic parameters of the  actin filament's growth. Different mechanisms of ATP hydrolysis in actin filaments  are compared. The possibility of adding or deleting oligomeric actin subunits is also  discussed. Finally, we  analyze the latest experiments on the growth dynamics of single actin filaments  \citep{fujiwara02, kuhn05}.

This article is organized as follows. The model of the actin polymerization is presented in the Section II. The effect of different mechanisms of hydrolysis on the actin filament growth is discussed in the Section III. The possibility of assembly and disassembly of oligomers is critically evaluated in the Section IV. The summary and conclusions are given in the Section V, while the technical details of calculations can be found in the Appendix.

\section*{Model of actin filament assembly}

Let us consider an actin filament as a two-stranded polymer, as shown in Fig. 1. It consists of two linear protofilaments. The size of the monomer subunit in this polymer is equal to $d=5.4$ nm, and  two protofilaments are shifted with respect to each other by a distance $a=d/2=2.7$ nm \citep{howard_book,bray_book}. Each monomer in the actin filament lattice carries a nucleotide molecule: it can be either ATP, or ADP (see Fig. 1). Shortly after actin monomers assemble into filaments, the ATP is hydrolyzed to ADP. For simplicity, we neglect the intermediate states of hydrolysis for actin monomers when  both products of hydrolysis, ADP and P$_{i}$ (inorganic phosphate), are bound to the actin monomer. Thus, we only consider two states of actin monomers in the filament - hydrolyzed and unhydrolyzed.  It is argued that only the dynamics of capped (unhydrolyzed) and uncapped (hydrolyzed) states effect the length fluctuations in the actin filaments \citep{vavylonis05}.  The hydrolyzed nucleotide remains bound to the polymer and at the physiological conditions it is not exchangeable with free ATP molecules from the solution. The dynamical and biochemical properties of ATP-bound (T-state) and ADP-bound (D-state) monomers are known to be different \citep{pollard86}. The dissociation rate of ADP-actin subunits from the actin filaments is estimated to be 2-5 times larger than the rate for the ATP-actin subunits, whereas the association rate is considerably slower (by a factor of 10) than that for the non-hydrolyzed analog \citep{pollard86, kuhn05}.

ATP hydrolysis plays an important role for the overall actin filament assembly dynamics, however the details of this process are not clear. Several mechanisms of ATP-actin hydrolysis in the filament have been proposed. In a random mechanism \citep{ohm94, pieper96, blanchoin02} any ATP-actin subunit can hydrolyze in a stochastic manner independently of the states of the neighboring monomers. The rate of hydrolysis in this case is proportional to the amount of non-transformed nucleotide in the polymer. A different approach is a sequential, or vectorial, mechanism \citep{carlier87, korn87, pantaloni85}, that assumes a high degree of cooperation  during the hydrolysis. According to this mechanism, the recently assembled actin monomer hydrolyzes its ATP only if it touches the more interior already hydrolyzed subunits. In this mechanism, there is a sharp boundary between the unhydrolyzed cap and the hydrolyzed core of the filament, while in the random mechanisms there are many interfaces between ATP-actin and ADP-actin subunits. Finally, it is also possible that a ``mixed''  mechanism, that combines the properties of random and vectorial approaches, describes the hydrolysis  in actin filaments. The available experimental data cannot clearly distinguish between these mechanisms. In our model we assume a vectorial mechanism, although, as we show below, the exact details of hydrolysis do not influence much the dynamic properties of the actin filament's growth.

There are infinite number of possible polymer configurations depending on the nucleotide state of each monomer and the geometry of polymer ends \citep{stukalin05}: see Fig. 1. However, we assume that only the so-called ``one-layer'' configurations, where the distance between two edge monomers at parallel protofilaments is less than $d$, are relevant for actin polymerization dynamics. This is based on the previous theoretical studies  \citep{stukalin04, stukalin05}, that showed that the one-layer approach is an excellent approximation to a full dynamic description of growth of two-stranded polymers with large lateral interactions between the subunits. It is also known that for actin filaments the lateral interaction energy is larger than 5 k$_{B}$T per monomer \citep{erickson89, stukalin05}, and it strongly supports the ``one-layer'' approximation.

Each configuration we label with two pairs of integers, ($l_{1},k_{1};l_{2},k_{2}$), where $l_{i}$ is the total number of monomers (hydrolyzed and unhydrolyzed) in the $i$-th protofilament, while $k_{i}$ specifies the number of ATP-actin subunits in the same protofilament. For example, the configuration $A$ from Fig.1 is labeled as (2,1;3,2), while the configuration $B$ is described as (2,1;3,1). The polymerization dynamics at both ends of the actin filament is considered independently from each other.

As shown in Fig. 1, at each end free ATP-actin molecules from the solution can attach to the actin filament with the rate $u=k_{T}c$, where $k_{T}$ is the ATP-actin polymerization rate constant and $c$ is the concentration of free ATP-actin species in the solution. Because of the excess of free ATP molecules in the solution only ATP-actin monomers are added to the filament \citep{carlier87, korn87, pantaloni85}.   We also  assume  that the dissociation rates of actin monomers  depend on their nucleotide state, and only the leading subunits dissociate from the filament. Specifically, ATP-actin monomer may detach with the rate $w_{T}$, while the hydrolyzed subunit dissociates with the rate $w_{D}$: see Fig. 1. In addition, the sequential vectorial mechanism of hydrolysis is assumed, i.e., ATP-actin monomer can transform into ADP-actin state with the rate $r_{h}$ if it touches two already hydrolyzed subunits.

The growth dynamics of single actin filaments can be determined by solving  a set of Master equations for all possible  polymer configurations. The mathematical derivations and all details of calculations are given in Appendix. Here we only present the explicit expressions for the dynamics properties of actin filament growth  at stationary state. Specifically, the mean growth velocity is equal to
\begin{equation}\label{vel}
V = \frac{d}{2} \{ u - w_T q - w_D (1-q) \}, 
\end{equation}
and dispersion is given by
\begin{equation}\label{disp}
D = \frac{d^2}{8} \left \{ u + w_T q + w_D (1-q) + \frac{2(w_D - w_T)(u + w_D q)}{w_T + r_h} \right \}, 
\end{equation}
for $0 \le q \le 1$, where
\begin{equation}\label{fract}
q = \frac{u}{w_T + r_h}.
\end{equation}

The parameter $q$ plays a critical role for understanding mechanisms of  actin  growth dynamics. It  has a meaning of  probability that the system  is in  a ``capped'' state with $N_{cap} \ge 1$ ATP-actin monomers , i.e.,  it is a fraction of time that the actin filament can be found in any  configuration with at least one unhydrolyzed subunit. For example, in Fig. 1 the configurations $A$ and $B$ are capped (with  $ N_{cap}=3$ and 2, correspondingly), while the configuration $C$ is uncapped with $ N_{cap}=0$. The parameter $q$ increases linearly with the concentration  of free ATP-actin monomers because  of the relation $u = k_{T} c $. However, it cannot be larger than 1, and this observation leads to existence of a special transition point with the concentration
\begin{equation}\label{c_trans}
c' = \frac{w_T + r_h}{k_T}.
\end{equation}
Above the transition point we have $ q (c \ge c') = 1 $, and the probability to have a polymer configuration with $N_{cap}=0$ is zero and the unhydrolyzed ATP cap grows steadily with time. At large times, for $c \ge c'$ the average length of ATP cap is essentially infinite, while below the transition point ($c < c'$) this length is always finite. 

At the critical concentration for each end of the filament, by definition, the mean growth velocity  for this end vanishes. Using Eqs.~\ref{vel} and \ref{fract} it can be shown that
\begin{equation}\label{c_crit}
c_{crit}= \frac{w_{D}}{k_{T}} \frac{(w_T + r_h)}{(w_{D}+r_{h})}.
\end{equation}
An important observation is the fact the  critical concentration is always below the transition point, 
\begin{equation}\label{cc}
c_{crit} = \frac{w_D}{w_D + r_h} c'.
\end{equation}

Because at concentrations larger than the transition point the dissociation events of hydrolyzed actin monomers are absent, the explicit expressions for the mean growth velocity and dispersion in this case are given by
\begin{equation}\label{vel_disp1}
V_{0} = \frac{d}{2} ( u - w_T ), \quad  D_{0} = \frac{d^{2}}{8} ( u + w_T ).
\end{equation}  
The calculated mean growth velocity for the barbed end of the actin filament is shown in Fig. 2 for  parameters specified in Table 1. It can be seen that the velocity depends linearly on the concentration of free ATP-actin particles in the solution, although the slope changes at the transition point. This is in agreement with experimental observations on actin filament's growth \citep{carlier86}. However, the behavior of  dispersion is very different --- see Fig. 3. It also grows linearly with concentration in both regimes, but there is a discontinuity in dispersion at the transition point. From Eqs. (\ref{disp}) and (\ref{vel_disp1}) we obtain that the size of the jump is equal to
\begin{equation}\label{ratio}
\frac{D(c')}{D_{0}(c')}=1+ \frac{2(w_{D}-w_{T})(w_{T}+w_{D}+r_{h})}{(w_{T}+r_{h})(2w_{T}+r_{h})}.
\end{equation}
The origin of this phenomenon is the fact that ADP-actin subunits dissociations contribute to overall growth dynamics only below the transition point $c'$. Note, as shown in Fig. 3, this contribution can increase the length fluctuations when $w_{D} > w_{T}$  (the  barbed end of the filament), or dispersion can be reduced  for $w_{D} < w_{T}$  (the pointed end of the filament). The jump disappears when $w_{D}=w_{T}$.  For the barbed end of the actin filament we calculate, using the parameters from the Table 1,  that $D(c')/D_{0}(c') \simeq 20.6$ and it approaches to 26.5 when $r_{h} \rightarrow 0$. This result agrees quite well with the experimentally observed ``apparent'' difference in the kinetic rate constants (35-40 times)  \citep{fujiwara02,kuhn05}.

\begin{table}
\begin{tabular}{l@{\hspace{0.7cm}}l@{\hspace{0.7cm}}c@{\hspace{0.7cm}}c@{\hspace{0.7cm}}l}
\multicolumn{4}{l}{\bfseries TABLE 1 \hspace{0.5cm} Summary of rate constants} \\ \hline
Rate constant & Reaction & Barbed end & Pointed end & Source \\ \hline
$k_{T}$, $\mu$M$^{-1}$ s$^{-1}$ & ATP-actin association  & 11.6 & 1.3 & Refs. \citep{pollard86, pollard00} \\ 
$w_{T}$, s$^{-1}$ & ATP-actin dissociation & 1.4 & 0.8 & 
Refs. \citep{pollard86, pollard00} \\
$w_{D}$, s$^{-1}$ & ADP-actin dissociation & 7.2 & 0.27 & 
Refs. \citep{pollard86, pollard00} \\
$r_h$, s$^{-1}$ & ATP hydrolysis & 0.3 & 0.3 & Ref. \citep{blanchoin02} \\ \hline
\end{tabular}
\end{table}

In order to compare our theoretical predictions with experimental observations the dynamics at both ends should be accounted for. However, as we showed earlier \citep{stukalin05}, the total  velocity of growth and the overall dispersion are the sums of the corresponding contributions for each end of the filament. The parameters we use in the calculations are shown in Table 1. Experimental measurements of actin filament's growth  suggest  that the hydrolysis rate $r_{h}$ is rather small, and we took $r_{h}=0.3$ s$^{-1}$ as given in the most recent investigation \citep{blanchoin02}. Note, however, that this value is determined assuming the random mechanism of hydrolysis. But the specific value of $r_{h}$ does not strongly influence our calculations. More controversial is the value of ADP-actin dissociation rate constant $w_{D}$. Most experiments indicate that $w_{D}$ is relatively large, ranging from  4.3 s$^{-1}$ \citep{kinosian02} to 11.5 s$^{-1}$ \citep{teubner98}, however the latest measurements performed using FRET method \citep{kuhn05} estimated that the dissociation rate is lower, $w_{D}=1.3$ s$^{-1}$. We choose for $w_{D}$ the value of 7.2 s$^{-1}$ as better describing the majority of experimental work.

For the actin filament system with the parameters given in Table 1 we can calculate from Eq.~\ref{c_crit} that the critical concentration for the barbed end is $c_{crit} \simeq 0.141$  $\mu$M, while for the pointed end it is equal to $c_{crit} \simeq 0.401$  $\mu$M. However, the contribution of the pointed end processes to the overall growth dynamics  is very small. As a result, the treadmilling concentration, when the overall growth rate vanishes, is estimated as $c_{tm} \simeq 0.144$ $\mu$M, and it is only slightly above the critical concentration for the barbed end (see Fig. 3). The treadmilling concentration also almost  coincides with the transition point for the barbed end, as can be calculated from Eq.~\ref{c_trans}, $c' \simeq 0.147$ $\mu$M. According to Eq.~\ref{disp}, the dispersion at treadmilling concentration at stationary-state conditions is equal to $D(c_{tm}) \simeq 31.6$ $sub^{2} \cdot$s$^{-1}$, with the contribution from the pointed end equal 0.5$\%$. From experiments, the values 29 $sub^{2} \cdot$s$^{-1}$ \citep{fujiwara02} and 31 $sub^{2} \cdot $s$^{-1}$ \citep{kuhn05} are reported for the filaments grown from Mg-ATP-actin monomers, and 25 $sub^{2} \cdot$s$^{-1}$ \citep{fujiwara02} is the dispersion for Ca-ATP-actin filaments. The agreement between theoretical predictions and experimental values is very good. It is also important to note that, in contrast to the previous theoretical description \citep{vavylonis05}, our  model predicts large length fluctuations slightly above the $c_{crit}$ for the barbed end of the filaments, exactly  as was observed in the experiments \citep{fujiwara02,kuhn05}. 

The presented theoretical model allows to calculate explicitly not only the dynamic properties of actin  growth but also the nucleotide composition of the filaments. As shown in Appendix, the mean size of the cap of  ATP-actin monomers is given by 
\begin{equation}\label{cap_size}
<N_{cap}> = \frac{q}{1-q}=\frac{c}{c'-c}.
\end{equation}
Then at the critical concentration for the barbed end, $c_{crit} \simeq 0.141$  $\mu$M, the cap size at the barbed end is $N_{cap} \simeq 24$, while the cap size at the pointed  end  at this concentration  (with the transition point $c' \simeq 0.846$ $\mu$M) is less than 1 monomer.  These results agree with the Monte Carlo computer simulations of actin polymerization dynamics \citep{vavylonis05}. Large ATP cap appears at the barbed end of actin filament and smaller cap is found  at the pointed end \citep{dufort93}. This is quite reasonable since the transition point for the barbed end  is much smaller than the corresponding one for the pointed end. The overall dependence of $N_{cap}$ on the concentration of free actin monomers is shown in Fig. 4.

\section*{Random vs. vectorial ATP hydrolysis in actin filaments.}

An important issue for understanding the actin polymerization dynamics is the nature of ATP hydrolysis mechanism. In our theoretical model the vectorial mechanism is utilized. In order to understand what features of actin dynamics are independent of the details of hydrolysis, it is necessary to compare the random and the vectorial mechanisms for this process. In our model of the polymer's  growth with vectorial mechanism, the actin filament consists of two parallel linear chains shifted by the distance $a=d/2$ from each other. According to our dynamic rules, the addition (removal) of one actin subunit increases (decreases) the overall length of the filament by the distance $a$. Then the growth dynamics of two-stranded polymers can be effectively mapped into the polymerization of single-stranded chains with an effective monomer's size $d_{eff}=d/2$.  

A single-stranded model of the actin filament's growth that assumes association of ATP-actin monomers and dissociation of ATP-actin and ADP-actin subunits  along with the random hydrolysis has been developed earlier \citep{keiser86,bindschadler04}. In this model the parameter $q$ is also introduced, and it has a meaning of the probability to find the leading subunit of the polymer in the unhydrolyzed state. However, the parameter  $q$ in the random hydrolysis model has a more complicated dependence on the concentration than in the vectorial model. It can be found as a root of the cubic equation,
\begin{equation}\label{q_rand}
\left[u - (u + w_T + r_h) q \right] (u - w_T q) + \left[ w_T q + w_D (1 - q)\right] \left[u - (w_T + r_h) q \right]  q = 0, 
\end{equation}
with the obvious restriction that $0 \le q \le 1$. 

Using the parameters given in the Table 1, the fraction of the capped configurations for two different mechanisms of hydrolysis is shown in Fig. 5. The predictions for both mechanisms are close at very low and very high concentrations, but deviate near  the critical concentration. It can be understood if we analyze  the special case  $w_{T}=w_{D}$, although all arguments are still valid in the general case of $w_{D} \ne w_{T}$. From Eq.~\ref{q_rand} we obtain
\begin{equation}
q=\frac{u+w_{T}+r_{h}}{2 w_{T}} \left( 1-\sqrt{1-\frac{4 w_{T} u}{(u+w_{T}+r_{h})^{2}}} \right).
\end{equation}
In the limit of very low concentrations the fraction $q$ approaches 
\begin{equation}
q \simeq \frac{u}{w_{T}+r_{h}},
\end{equation}
while for $c \gg 1$ it can be described as
\begin{equation}
q \simeq 1- \frac{w_{T}+r_{h}}{u}.
\end{equation}
 Generally, in the limit of very low hydrolysis rates the random and the vectorial mechanisms should predict the same dynamics, as expected. It can be seen by taking the limit of $r_{h} \rightarrow 0$ in Eq.~\ref{q_rand}, which gives $q=u/w_{T}$ for $u < w_{T}$ and $q=1$ for $u > w_{T}$. These results are illustrated in Fig. 5.

In order to calculate the size of the ATP-cap in the actin filament for the random mechanism we introduce a function $P_{n}$ defined as a probability to find in the ATP-state the monomer positioned  $n$ subunits away from the leading one. Then it can be shown that this probability is exponentially decreasing  function of $n$ \citep{keiser86},
\begin{equation}
\frac{P_{n+1}}{P_{n}}= 1- \frac{r_{h} q}{u-w_{T} q}, \quad P_{1}=q.
\end{equation} 
The size of the unhydrolyzed cap in the polymer is associated with the total number of ATP-actin monomers \citep{vavylonis05},
\begin{equation}
N_{cap}=\sum_{n=1}^{\infty} P_{n}= \frac{u-w_{T} q}{r_{h}}.
\end{equation} 
The results of the different mechanisms for  $N_{cap}$ are plotted in Fig. 4. The random and vectorial mechanism agree at low concentrations, but the predictions differ for large concentrations. Only in the limit of small hydrolysis rates the predictions from the two hydrolysis mechanisms start to converge. 

The mean growth velocity in the model with the random mechanism is given by
\begin{equation}\label{vel_rand}
V = d_{eff} \left [ u - w_T q - w_D (1-q) \right ], 
\end{equation}
which is exactly the same as in the vectorial mechanism (see Eq.~\ref{vel}), although the fraction of ATP-cap configurations $q$ have a different behavior in  two models. Mean growth velocities for different mechanisms are compared in Fig. 2. Again, the predictions for different mechanisms of hydrolysis converge for very low and very high concentrations of free actin, but differences arise  near the critical concentration.

In the model with the random mechanism of hydrolysis  \citep{keiser86,vavylonis05} the analytical expressions for  dispersion have not been found. However, from the comparison of the fraction of ATP-capped configurations $q$, the size of the ATP-cap $N_{cap}$ and the mean growth velocity $V$ it can be concluded that both mechanisms predict qualitatively and quantitatively similar picture for the dynamic behavior of the single actin filaments. Thus, it can be expected that, similarly to the vectorial mechanism,  there is a peak in dispersion near the critical concentration in the model with  the random hydrolysis, in agreement with the latest Monte Carlo computer simulations results \citep{vavylonis05}.

\section*{Assembly/disassembly of oligomers in actin filament dynamics}

The association and dissociation of large oligomers of actin monomers  has been suggested as a possible reason for large fluctuations during the elongation of single actin filaments  \citep{fujiwara02, kuhn05}. Let us consider this possibility more carefully. Suppose that the oligomeric particles that contain $n$ ATP-actin monomers can attach to or detach from the filament. Then the mean growth velocity can be written as
\begin{equation}\label{vel_oligo}
V(n) = \frac{nd}{2}[u(n) - w_T(n)],
\end{equation}
where $u(n)$ and $w_{T}(n)$ are the  assembly and disassembly rates of oligomeric subunits. Similarly, the expression for dispersion is given by
\begin{equation}\label{disp_oligo}
D(n) =  \frac{(nd)^2}{8}\left[ u(n) + w_T(n)\right ]. 
\end{equation}

At the same time, in the analysis of the experimental data \citep{fujiwara02, kuhn05} the addition or removal of single subunits has been assumed. It means that the rates has been measured using the following expression,
\begin{equation}
V = \frac{d}{2}[u^{eff} - w_T^{eff}].
\end{equation}
Comparing this equation with Eq.~\ref{vel_oligo}, it yields the relation between the effective rates $u^{eff}$ and $w_T^{eff}$ per monomer and the actual rates $u(n)$ and $w_T(n)$ per oligomer,\begin{equation}
u^{eff}=nu(n), \quad w_T^{eff}=n w_T(n).
\end{equation}
The substitution of these effective rates into the expression for dispersion (\ref{disp_oligo}) with $n=1$ produces
\begin{equation}
D(n=1) =  \frac{d^2}{8}\left[ u^{eff} + w_T^{eff} \right ]= \frac{nd^2}{8}\left[ u(n) + w_T(n)\right ]=\frac{D(n)}{n}. 
\end{equation}
It means that dispersion calculated assuming the association/dissociation of monomers underestimates the ``real'' dispersion in $n$ times, but not in $n^{2}$ times as was suggested earlier \citep{fujiwara02, kuhn05}.

The experimental results \citep{fujiwara02, kuhn05} suggest that only the addition or dissociation of oligomers with $n=35-40$ can explain the large length fluctuation in the single actin filaments if one accepts the association/dissociation of oligomers. These particles are quite large by size ($\simeq$100 nm), and, if present in the system, they would be easily observed in the experiments. However,  no detectable amounts of large oligomers have been found  in studies of kinetics of the actin polymerization. It has been reported \citep{attri91} that only small oligomers (up to $n=4-8$)  may coexist with the monomers and the polymerized actin under the {\it special} (not physiological) solution conditions. Therefore, it is very unlikely, that the presence of very small (if any) amounts of such oligomers might influence the dynamics and large length fluctuations in the actin growth.

\section*{Summary and conclusions}

The growth dynamics of single actin filaments is investigated theoretically   using the stochastic model that takes into account the dynamics at both ends of filament, the structure of the polymer's tip, lateral interactions between the protofilaments,  the hydrolysis of ATP bounded to the actin subunit, and assembly and disassembly of hydrolyzed and unhydrolyzed actin monomers. It is assumed that sequential (vectorial) mechanism of hydrolysis controls the transformation of ATP-actin subunits. Using the analytical approach,  exact expressions for the mean growth velocity, the length dispersion, and the mean size of fluctuating ATP-cap are obtained in terms of the  kinetic rate constants that describe the assembly and disassembly events, and the hydrolysis of nucleotides. It is shown that there are two regimes of single actin filament's growth. At high concentrations the size of the ATP-cap is very large and the the fully hydrolyzed core is never exposed at filament's tip. As a result, the disassembly of ADP-actin subunits does not contribute to the overall dynamics. The situation is different at low concentrations, where the size of ATP-cap is always finite. Here the dissociation of both hydrolyzed and unhydrolyzed  actin monomers is critical for the growth dynamics of filaments. The boundary between two regimes is defined by the transition point, that depends on the association/dissociation rate constants for the ATP-actin monomers and on the hydrolysis rate. For any non-zero rate of hydrolysis the transition point is {\it always} above the critical concentration where the mean growth velocity becomes equal to zero. The most remarkable result of our theoretical analysis is the non-monotonous behavior of dispersion as the function of concentration and large length fluctuations near the critical concentration. These large fluctuations are explained by alternation between the relatively slow assembly/disassembly of ATP-actin subunits and the rapid dissociation of ADP-actin monomers.

For the experimentally determined kinetic rates our theoretical analysis suggests that the contribution of the dynamics at the pointed end of the actin filament is very small. The treadmilling concentration for the system, as well as the transition point for the barbed end, are only slightly above the critical concentration for the barbed end. At this transition point the dispersion of length reaches a maximal value, and it is smaller for larger concentrations. Our theoretical predictions are in excellent agreement with all available experimental observations on the dynamics of single actin filaments. However, more measurements of dispersion and other dynamic properties at different monomeric concentrations are needed in order to check the validity of the presented theoretical method.

Since the exact mechanism of ATP hydrolysis in actin is not known, we discussed and compared the random and the vectorial mechanisms for the simplified effective single-stranded model of actin growth. It was shown that the mean growth velocity, the fraction of the configurations with unhydrolyzed cap and the size of ATP-cap are qualitatively and quantitatively similar at low and at high concentrations, although there are deviations near the transition point. It was suggested then that the length fluctuations in the random mechanism, like in the vectorial mechanism, might also exhibit a peak near the critical concentration. Future experimental measurements of dynamic properties at different concentrations might help to distinguish between the hydrolysis mechanisms in the actin growth.    

The possibility of attachment and detachment of large oligomers in the single actin filaments has been also discussed. It was argued that the experimental observations of large length fluctuations can only be explained by addition of oligomers consisting of 35-40 monomers. However, these oligomers have a large size and such events have not been observed  or have been excluded from the analysis of the experiments. Thus the effect of the oligomer assembly/disassembly on the single actin filament's growth is probably negligible.

Although the effect of ATP hydrolysis on polymerization dynamics of actin filaments has been studied studied before \citep{keiser86,hill86, hill87,vavylonis05}, to best of our knowledge,  the present work is the first that provides rigorous calculations of the mean growth velocity, dispersion, the size of ATP-cap and the fraction of capped configurations simultaneously.   It is reasonable to suggest that this method might be used to investigate the dynamic instability in microtubules because the polymer can be viewed as growing in two dynamic phases. In one phase the  ATP-cap is always  present at the end of the filament, while in the second phase it is absent. Similar approach to investigate the dynamic phase changes has been proposed earlier \citep{hill86}.  The model can also be improved by  considering  the intermediate states of hydrolysis and the release  of inorganic phosphate  \citep{pollard03,vavylonis05}, and the possible exchange of nucleotide at the terminal subunit of the barbed end of the actin filaments \citep{teubner98}.

\section*{Acknowledgments}

The authors would like to acknowledge the support from the Welch Foundation (grant  C-1559), the Alfred P. Sloan foundation (grant  BR-4418)  and the U.S. National Science Foundation (grant CHE-0237105). The authors also are grateful to M.E. Fisher for valuable comments and discussions.

\section*{Appendix: One-layer polymerization model with sequential ATP hydrolysis for two-stranded polymers.}

\setcounter{equation}{0}

\renewcommand{\theequation}{\mbox{A\arabic{equation}}}

Let us define a function $P(l_{1},k_{1};l_{2},k_{2};t)$ as the probability of finding the two-stranded polymer in the configuration  $(l_{1},k_{1};l_{2},k_{2})$. Here  $l_{i}, k_{i} = 0, 1,...$ ($k_{i} \le l_{i}$, $i=1$ or 2) are two independent parameters that count  the total number of subunits ($l_{i}$) and the number of unhydrolyzed subunits ($k_{i}$) in the $i$-th protofilament. We assume that the polymerization and hydrolysis in the actin filament can be described by  ``one-layer'' approach \citep{stukalin05,stukalin04}. It means that $l_{2}=l_{1}$ or $l_{2}=l_{1}+1$, and $k_{2}=k_{1}$ or $k_{2}=k_{1} \pm 1$ (see Fig. 1).  Then the probabilities can be described by a set of  master equations. For configurations with $l_{1}=l_{2}=l$ and $1 \le k < l$ we have 
\begin{eqnarray}\label{me1}
\frac{dP(l,k;l,k;t)}{{dt}} =  u P(l - 1,k-1;l,k;t) + w_T  P(l,k;l + 1,k+1;t)  \nonumber \\
  + r_h P(l,k+1;l,k;t) - (u + w_T + r_h )P(l,k;l,k;t), &  
\end{eqnarray}
and
\begin{eqnarray}\label{me3}
 \frac{dP(l,k;l,k-1;t)}{{dt}} =  u P(l - 1,k-1;l,k-1;t) + w_T  P(l,k;l + 1,k;t) \nonumber \\
 + r_h P(l,k;l,k;t) - (u + w_T + r_h )P(l,k;l,k-1;t). &
\end{eqnarray}
Similarly for the configurations with $l_{1}=l_{2}-1=l$ and  $1 \le k < l+1$ the master equations are
\begin{eqnarray}\label{me2}
\frac{dP(l,k-1;l + 1,k;t)}{dt} =  u P(l,k-1;l,k-1;t) + w_T P(l + 1,k;l + 1,k;t)  \nonumber \\
 + r_h P(l,k;l + 1,k;t) - (u + w_T + r_h )P(l,k-1;l + 1,k;t), &  
\end{eqnarray}
and
\begin{eqnarray}\label{me4}
\frac{dP(l,k;l + 1,k;t)}{dt} =   u  P(l,k;l,k-1;t) + w_T P(l + 1,k+1;l + 1,k;t)  \nonumber \\
 + r_h P(l,k;l + 1,k+1;t) - (u  + w_T  + r_h )P(l,k;l + 1,k;t). &
\end{eqnarray} 
Then the polymer configurations without ATP-actin monomers ($k=0$) can be described by
\begin{eqnarray}\label{me5}
\frac{dP(l,0;l,0;t)}{{dt}} = & w_T  P(l,0;l + 1,1;t) + w_D  P(l,0;l + 1,0;t) & \nonumber \\
& - (u + w_D )P(l,0;l,0;t), &  
\end{eqnarray}
and 
\begin{eqnarray}\label{me6}
\frac{dP(l,0;l+1,0;t)}{{dt}} = &  w_T  P(l+1,1;l + 1,0;t) + w_D P(l+1,0;l + 1,0;t) & \nonumber \\
& -(u + w_D )P(l,0;l+1,0;t). &
\end{eqnarray}
Finally, for the configurations consisting of  only unhydrolyzed subunits  we have
\begin{eqnarray}\label{me7}
\frac{dP(l,l;l,l;t)}{{dt}} =  u P(l - 1,l-1;l,l;t) + w_T  P(l,l;l + 1,l+1;t)  \nonumber \\
  - (u + w_T + r_h )P(l,l;l,l;t), &  
\end{eqnarray}
and
\begin{eqnarray}\label{me8}
\frac{dP(l,l;l + 1,l+1;t)}{dt} =  u  P(l,l;l,l;t) + w_T P(l + 1,l+1;l + 1,l+1;t)  \nonumber \\
  - (u  + w_T   + r_h )P(l,l;l + 1,l+1;t), &  
\end{eqnarray}

The conservation of probability leads to  
\begin{eqnarray}
\sum\limits_{k = 0}^{ + \infty } \left[ \sum\limits_{l = 0}^{ + \infty } P(l,k;l,k;t) + \sum\limits_{l = 0}^{ + \infty } P(l,k+1;l,k;t) \right. \nonumber \\
 \left. + \sum\limits_{l = 0}^{ + \infty } P(l,k;l+1,k+1;t) + \sum\limits_{l = 0}^{ + \infty } P(l,k;l+1,k;t) \right]  = 1, 
\end{eqnarray}
at all times.

Following the method of Derrida \citep{derrida83}, we define two sets of auxiliary functions ($k = 0,1,...$),
\begin{eqnarray}
& &  B_{k,k}^0 (t) = \sum\limits_{l \not= k}^{ + \infty } {P(l,k;l,k;t)}, \nonumber \\
& &  B_{k+1,k}^0 (t)= \sum\limits_{l = 0}^{ + \infty } {P(l,k+1;l, k;t)}, \nonumber \\
& &  B_{k,k+1}^1 (t)= \sum\limits_{l \not= k}^{ + \infty } {P(l,k;l+1,k+1;t)},  \nonumber \\
& &  B_{k,k}^1(t)= \sum\limits_{l = 0}^{ + \infty } {P(l,k;l+1, k;t)}, \nonumber \\
& &  B^0 (t)= \sum\limits_{l = 0}^{ + \infty } {P(l,l;l,l;t)},  \nonumber \\
& &  B^1 (t)= \sum\limits_{l = 0}^{ + \infty } {P(l,l;l+1,l+1;t)};
\end{eqnarray} 
and 
\begin{eqnarray}
& &  C_{k,k}^0 (t) = \sum\limits_{l \not= k}^{ + \infty } {(l + \frac{1}{2}) P(l,k;l,k;t)}, \nonumber \\
& &  C_{k+1,k}^0 (t) = \sum\limits_{l = 0}^{ + \infty } {(l + \frac{1}{2}) P(l,k+1;l,k;t),} \nonumber \\
& &  C_{k,k+1}^1 (t)= \sum\limits_{l \not= k}^{ + \infty } {(l + 1 ) P(l,k;l+1,k+1;t)}, \nonumber \\
& &  C_{k,k}^1 (t)= \sum\limits_{l = 0}^{ + \infty } {(l + 1 ) P(l,k;l+1,k;t)}, \nonumber \\
& &  C^0 (t)= \sum\limits_{l = 0}^{ + \infty } {(l + \frac{1}{2}) P(l,l;l,l;t)},  \nonumber \\
& &  C^1 (t)= \sum\limits_{l = 0}^{ + \infty } {(l + 1 ) P(l,l;l+1,l+1;t)}.
\end{eqnarray}
Note that the conservation of probability gives us
\begin{equation}\label{norm}
\sum\limits_{k = 0}^{ + \infty } {B_{k,k}^0 (t) + B_{k+1,k}^0 (t) + B_{k,k+1}^1 (t) + B_{k,k}^1 (t) + B^0 (t) + B^1 (t)} = 1.
\end{equation}
Then from the master equations (\ref{me1}), (\ref{me3}), (\ref{me2}) and  (\ref{me4}) we derive  for $k \ge 1$
\begin{eqnarray}\label{Be1}
  \frac{dB_{k,k}^0 (t)}{dt} = & u B_{k-1,k}^1 (t) + w_T B_{k,k+1}^1 (t) + r_h B_{k+1,k}^0(t) - (u + w_T + r_h) B_{k,k}^0(t), & \nonumber \\
 \frac{dB_{k-1,k}^1 (t)}{dt} = & u B_{k-1,k-1}^0 (t) + w_T B_{k,k}^0(t) + r_h B_{k,k}^1 (t) - (u + w_T + r_h) B_{k-1,k}^1 (t), &   
\nonumber \\
\frac{dB_{k,k-1}^0 (t)}{dt} = & u B_{k-1,k-1}^1 (t) + w_T B_{k,k}^1 (t) + r_h B_{k,k}^0 (t) - (u + w_T + r_h) B_{k,k-1}^0 (t), & 
\nonumber \\
\frac{dB_{k,k}^1 (t)}{dt} = & u B_{k,k-1}^0 (t) + w_T B_{k+1,k}^0 (t) + r_h B_{k,k+1}^1 (t) - (u + w_T + r_h) B_{k,k}^1 (t), & 
\end{eqnarray} 
while the master equations (\ref{me5}), (\ref{me6}) for $k = 0$ yield 
\begin{eqnarray}\label{Be2}
\frac{dB_{0,0}^0 (t)}{dt} = & w_T B_{0,1}^1(t) + w_D B_{0,0}^1 (t) + r_h B_{1,0}^0 (t) - (u + w_D) B_{0,0}^0 (t), & \nonumber \\
\frac{dB_{0,0}^1 (t)}{dt} = & w_T B_{1,0}^0 (t) + w_D B_{0,0}^0 (t) + r_h B_{0,1}^1 (t) - (u + w_D) B_{0,0}^1 (t). &
\end{eqnarray}
Finally, equations (\ref{me7}) and (\ref{me8}) lead to
\begin{eqnarray}\label{Be3}
\frac{dB^0 (t)}{dt} = & (u + w_T) B^1 (t) - (u + w_T + r_h) B^0 (t), & \nonumber \\
\frac{dB^1 (t)}{dt} = & (u + w_T) B^0 (t) - (u + w_T + r_h) B_1 (t), & 
\end{eqnarray}

Similar arguments can be used to describe the functions $C_{k,k}^0$, $C_{k+1,k}^0 $, $C_{k,k+1}^1$, and $C_{k,k}^1$. Specifically, for $k \ge 1$ we obtain
\begin{eqnarray}\label{Ce1}
\frac{dC_{k,k}^0 (t)}{dt} = & u C_{k-1,k}^1 (t) + w_T C_{k,k+1}^1 (t) + r_h C_{k+1,k}^0 (t) - (u + w_T + r_h )C_{k,k}^0 (t) & \nonumber \\
& + \frac{1}{2}[u B_{k-1,k}^1 (t) - w_T B_{k,k+1}^1 (t)], &
\end{eqnarray}
\begin{eqnarray}\label{Ce2}
\frac{dC_{k-1,k}^1 (t)}{dt} = & u C_{k-1,k-1}^0 (t) + w_T C_{k,k}^0 (t) + r_h C_{k,k}^1 (t) - (u + w_T + r_h )C_{k-1,k}^1 (t) & \nonumber \\
& + \frac{1}{2} [u B_{k-1,k-1}^0 (t) - w_T B_{k,k}^0 (t)], &
\end{eqnarray} 
\begin{eqnarray}\label{Ce3}
\frac{dC_{k,k-1}^0 (t)}{dt} = & u C_{k-1,k-1}^1 (t) + w_T C_{k,k}^1 (t) + r_h C_{k,k}^0 (t) - (u + w_T + r_h )C_{k,k-1}^0 (t) & \nonumber \\
& + \frac{1}{2} [u B_{k-1,k-1}^1 (t) - w_T B_{k,k}^1 (t)], &
\end{eqnarray}
\begin{eqnarray}\label{Ce4}
\frac{dC_{k,k}^1 (t)}{dt} = & u C_{k,k-1}^0 (t) + w_T C_{k+1,k}^0 (t) + r_h C_{k,k+1}^1 (t) - (u + w_T + r_h )C_{k,k}^1 (t) & \nonumber \\
& + \frac{1}{2} [u B_{k,k-1}^0 (t) - w_T B_{k+1,k}^0 (t)]. &
\end{eqnarray}
For $k = 0$ the expressions are the following,
\begin{eqnarray}\label{Ce5}
\frac{dC_{0,0}^0 (t)}{dt} = & w_T C_{0,1}^1 (t) + w_D C_{0,0}^1(t) + r_h C_{1,0}^0(t) - (u + w_D)C_{0,0}^0 (t) & \nonumber \\
& - \frac{1}{2}[w_T B_{0,1}^1 (t) + w_D B_{0,0}^1(t)], &
\end{eqnarray}
\begin{eqnarray}\label{Ce6}
\frac{dC_{0,0}^1 (t)}{dt} = & w_T C_{1,0}^0 (t) + w_D C_{0,0}^0(t) + r_h C_{0,1}^1 (t) - (u + w_D)C_{0,0}^1 (t) & \nonumber \\
& - \frac{1}{2}[w_T B_{1,0}^0 (t) + w_D B_{0,0}^0(t)]. &
\end{eqnarray}

Again following the Derrida's approach \citep{derrida83} we introduce an ansatz that should be valid at large times $t$, namely,
\begin{equation}\label{ansatz}
B_{k,m}^i (t) \to b_{k,m}^i ,{\text{ }}C_{k,m}^i (t) \to a_{k,m}^i t + T_{k,m}^i \hspace{0.5cm} (i = 0,1; |k-m| \le 1).
\end{equation}
At steady state $dB_{k,m}^i (t)/dt = 0$, and Eqs.~\ref{Be1} and \ref{Be2}  yield for $k \ge 1$
\begin{eqnarray}\label{be1}
& 0 = u b_{k-1,k}^1 + w_T b_{k,k+1}^1 + r_h b_{k+1,k}^0 - (u + w_T + r_h) b_{k,k}^0, & \nonumber \\
& 0 = u b_{k-1,k-1}^0 + w_T b_{k,k}^0 + r_h b_{k,k}^1 - (u + w_T + r_h) b_{k-1,k}^1 , &  \nonumber \\
& 0 = u b_{k-1,k-1}^1 + w_T b_{k,k}^1 + r_h b_{k,k}^0 - (u + w_T + r_h) b_{k,k-1}^0 , & \nonumber \\
& 0 = u b_{k,k-1}^0 + w_T b_{k+1,k}^0 + r_h b_{k,k+1}^1 - (u + w_T + r_h) b_{k,k}^1 , & 
\end{eqnarray} 
while for $k = 0$ we obtain
\begin{eqnarray}\label{be2}
& 0 = w_T b_{0,1}^1 + w_D b_{0,0}^1 + r_h b_{1,0}^0 - (u + w_D) b_{0,0}^0, & \nonumber \\
& 0 = w_T b_{1,0}^0 + w_D b_{0,0}^0 + r_h b_{0,1}^1 - (u + w_D) b_{0,0}^1 . &
\end{eqnarray}
Finally, from Eq.~\ref{Be3} we have
\begin{eqnarray}\label{be3}
& 0 = (u + w_T) b^1 - (u + w_T + r_h) b^0, & \nonumber \\
& 0 = (u + w_T) b^0 - (u + w_T + r_h) b_1. & 
\end{eqnarray}

Due to the symmetry of the system  we can conclude that the probabilities $b_{k,k}^0=b_{k,k}^1$ and $b_{k+1,k}^0=b_{k,k+1}^1$. Then the solutions of Eqs.~\ref{be1} and \ref{be2} can be written in the following form,
\begin{eqnarray}\label{b_sol}
& b_{k,k}^0 = b_{k,k}^1 = \frac{1}{2}(1-q)q^{2k}, & \nonumber \\
& b_{k+1,k}^0  =  b_{k,k+1}^1 = \frac{1}{2}(1-q)q^{2k+1}, &  
\end{eqnarray}
where  $k = 0,1,..$, and
\begin{equation}
q = \frac{u}{w_T + r_h} < 1.
\end{equation}
In addition, Eqs.~\ref{be3} have only a trivial solution $ b^0 = b^1 = 0 $. Recall that $b^0$ and $b^1$  give the stationary-state probabilities of the polymer configurations with all subunits in ATP state, i.e., it corresponds to the case of very large $k$.  The solution   agrees with the results for  $ b_{k,k}^0 $ and $ b_{k,k+1}^1$ at $k \rightarrow \infty $ (see Eqs.~\ref{b_sol}).

For $q > 1 $ the systems of equations (\ref{be1}) and (\ref{be2}) have the trivial solutions, with all  $b_{k,m}^i=0$ for finite $k$ and $ m$. It means that at the stationary conditions the polymer can only exist in the configurations with very large number of unhydrolyzed subunits and the size of ATP-cap is infinite, while for $q<1$ the size of ATP-cap is always finite. The case $q=1$ is a boundary between two regimes. At this condition there is a qualitative change in the  dynamic properties of the system.   

To determine the coefficients $a_{k,m}^i$ and $T_{k,m}^i$ from  Eq.~\ref{ansatz}, the ansatz for the functions $C_{k,m}^i $ is substituted into the asymptotic expressions (\ref{Ce1} - \ref{Ce6}), yielding for $k \ge 1 $,
\begin{eqnarray}\label{ae1}
& 0 = u a_{k-1,k}^1 + w_T a_{k,k+1}^1 + r_h a_{k+1,k}^0 - (u + w_T + r_h) a_{k,k}^0, & \nonumber \\
& 0 = u a_{k-1,k-1}^0 + w_T a_{k,k}^0 + r_h a_{k,k}^1 - (u + w_T + r_h) a_{k-1,k}^1 , &  \nonumber \\
& 0 = u a_{k-1,k-1}^1 + w_T a_{k,k}^1 + r_h a_{k,k}^0 - (u + w_T + r_h) a_{k,k-1}^0 , & \nonumber \\
& 0 = u a_{k,k-1}^0 + w_T a_{k+1,k}^0 + r_h a_{k,k+1}^1 - (u + w_T + r_h) a_{k,k}^1 , & 
\end{eqnarray} 
At the same time, for $k = 0$ we obtain
\begin{eqnarray}\label{ae2}
& 0 = w_T a_{0,1}^1 + w_D a_{0,0}^1 + r_h a_{1,0}^0 - (u + w_D) a_{0,0}^0, & \nonumber \\
& 0 = w_T a_{1,0}^0 + w_D a_{0,0}^0 + r_h a_{0,1}^1 - (u + w_D) a_{0,0}^1. &
\end{eqnarray}
The coefficients $T_{k,m}^i$ satisfy the following equations (for $k \ge 1$),
\begin{eqnarray}\label{te1}
a_{k,k}^0 = & u T_{k-1,k}^1 + w_T T_{k,k+1}^1 + r_h T_{k+1,k}^0 - (u + w_T + r_h )T_{k,k}^0 & \nonumber \\
& + \frac{1}{2}[u b_{k-1,k}^1  - w_T b_{k,k+1}^1], &
\end{eqnarray}
\begin{eqnarray}\label{te2}
a_{k-1,k}^1 = & u T_{k-1,k-1}^0 + w_T T_{k,k}^0 + r_h T_{k,k}^1 - (u + w_T + r_h )T_{k-1,k}^1 & \nonumber \\
& + \frac{1}{2} [u b_{k-1,k-1}^0 - w_T b_{k,k}^0 ], &
\end{eqnarray} 
\begin{eqnarray}\label{te3}
a_{k,k-1}^0 = & u T_{k-1,k-1}^1 + w_T T_{k,k}^1 + r_h T_{k,k}^0 - (u + w_T + r_h )T_{k,k-1}^0 & \nonumber \\
& + \frac{1}{2} [u b_{k-1,k-1}^1 - w_T b_{k,k}^1], &
\end{eqnarray}
\begin{eqnarray}\label{te4}
a_{k,k}^1  = & u T_{k,k-1}^0 + w_T T_{k+1,k}^0 + r_h T_{k,k+1}^1 - (u + w_T + r_h )T_{k,k}^1 & \nonumber \\
& + \frac{1}{2} [u b_{k,k-1}^0 - w_T b_{k+1,k}^0]. &
\end{eqnarray}
For $k = 0$ the expressions are given by
\begin{eqnarray}\label{te5}	
a_{0,0}^0 = & w_T T_{0,1}^1 + w_D T_{0,0}^1 + r_h T_{1,0}^0 - (u + w_D)T_{0,0}^0 & \nonumber \\
& - \frac{1}{2}[w_T b_{0,1}^1 + w_D b_{0,0}^1], &
\end{eqnarray}
\begin{eqnarray}\label{te6}
a_{0,0}^1 = & w_T T_{1,0}^0 + w_D T_{0,0}^0 + r_h T_{0,1}^1 - (u + w_D)T_{0,0}^1 & \nonumber \\
& - \frac{1}{2}[w_T b_{1,0}^0 + w_D b_{0,0}^0]. &
\end{eqnarray}

Comparing Eqs.~\ref{be1} and \ref{be2} with expressions \ref{ae1} and \ref{ae2}, we conclude that
\begin{equation}\label{eq_a}
a_{k,m}^i  = Ab_{k,m}^i, \hspace{0.5cm} (i = 0,1),
\end{equation}
with the constant $A$. This constant can be calculated by summing over the left and right sides in Eq.~\ref{eq_a} and recalling the  normalization condition (\ref{norm}). The summation over all  $a_{k,i}$ in Eqs.~\ref{ae1} and \ref{ae2} produces
\begin{eqnarray}\label{eq_A}
& A = \sum\limits_{k = 0}^{ + \infty } \left[ a_{k,k}^0 + a_{k+1,k}^0 + a_{k,k+1}^1 + a_{k,k}^1 \right]  & \nonumber \\
& = \frac{1}{2} \left[ (u - w_T) - (w_D - w_T)(b_{0,0}^0 + b_{0,0}^1) \right].
\end{eqnarray}

To determine the coefficients $T_{k,m}^i$ , we need to solve Eqs.~\ref{te1}-\ref{te6}. Again, due to the symmetry, we have $T_{k,k}^0 = T_{k,k}^1 \equiv T_{2k} $, and  $T_{k+1,k}^0 = T_{k,k+1}^1 \equiv T_{2k+1}$ for all $k$. The solutions for these equations are given by
\begin{equation}\label{T}
T_k = q^{k} T_0 + \frac{1}{4} \frac{(u + w_D q)(1 - q)}{w_T + r_h}kq^{k-1} , 
\end{equation}
where $k = 0,1,...$ and $T_0$ is an arbitrary constant.

It is now possible to calculate explicitly the mean growth velocity, $V$, and dispersion, $D$, at steady-state conditions. The average length of the polymer is given by
\begin{eqnarray}\label{eq_l}
< l(t)>  & =   d \left(  \sum\limits_{k = 0}^{+ \infty} \sum\limits_{l = 0}^{+ \infty} (l + \frac{1}{2})P(l,k;l,k;t) + \sum\limits_{k = 0}^{+ \infty} \sum\limits_{l = 0}^{+ \infty} (l + \frac{1}{2})P(l,k+1;l,k;t) \right. \nonumber \\ 
  & \left. \quad +  \sum\limits_{k = 0}^{+ \infty} \sum\limits_{l = 0}^{+ \infty} (l + 1)  P(l,k;l,k+1;t) + \sum\limits_{k = 0}^{+ \infty} \sum\limits_{l = 0}^{+ \infty} (l + 1)P(l,k;l+1,k;t) \right) \nonumber \\
  & = d\sum\limits_{k = 0}^{ + \infty } \left[ C_{k,k}^0 (t) + C_{k+1,k}^0 (t) + C_{k,k+1}^1 (t) + C_{k,k}^1 (t)\right].
\end{eqnarray}
Then, using Eq.~\ref{eq_a}, we obtain for the velocity
\begin{equation}
V =  \lim_{t \to \infty } \frac{d}{{dt}} < l(t) >  = dA\left( {\sum\limits_{k = 0}^{ + \infty } {b_{k,k}^0 }  + \sum\limits_{k = 0}^{ + \infty } {b_{k+1,k}^0 } + \sum\limits_{k = 0}^{ + \infty } {b_{k,k+1}^1 } + \sum\limits_{k = 0}^{ + \infty } {b_{k,k}^1 } } \right) = dA.
\end{equation}

A similar approach can be used to derive the expression for dispersion. We start from
\begin{eqnarray}
<l^2 (t)>  & =  d^2 \left(   \sum\limits_{k = 0}^{+ \infty} \sum\limits_{l = 0}^{+ \infty} (l + \frac{1}{2})^2 P(l,k;l,k;t) + \sum\limits_{k = 0}^{+ \infty} \sum\limits_{l = 0}^{+ \infty} (l + \frac{1}{2})^2 P(l,k+1;l,k;t) \nonumber  \right. \\ 
& \left.  \quad + \sum\limits_{k = 0}^{+ \infty} \sum\limits_{l = 0}^{+ \infty} (l + 1)^2  P(l,k;l,k+1;t) + \sum\limits_{k = 0}^{+ \infty} \sum\limits_{l = 0}^{+ \infty} (l + 1)^2 P(l,k;l+1,k;t) \right).
\end{eqnarray}
Then, using the master equations (\ref{me1} - \ref{me6}), it can be shown that
\begin{eqnarray}\label{eq_l1}
\lim_{t \to \infty } \frac{d}{dt} <l^2 (t)> & = d^2  \left [ (u - w_T) \sum\limits_{k = 0}^{ + \infty } \{C_{k,k}^0 + C_{k+1,k}^0 + C_{k,k+1}^1 + C_{k,k}^1 \} - (w_D - w_T)(C_{0,0}^0 + C_{0,0}^1) \right. \nonumber \\ 
& \left. + \frac{1}{4} (u + w_T) + \frac{1}{4} (w_D - w_T)(b_{0,0}^0 + b_{0,0}^1) \right ]. 
\end{eqnarray}
Also, the following equation can be derived using Eq.~\ref{eq_l},
\begin{equation}\label{eq_l2}
\lim_{t \to \infty } \frac{d}{{dt}} \left(<l(t)>^2\right)  =2 d^2 A\sum\limits_{k = 0}^{ + \infty } \left[ C_{k,k}^0  + C_{k+1,k}^0 + C_{k,k+1}^1  + C_{k,k}^1 \right]. 
\end{equation}
The formal expression for dispersion is given by
\begin{equation}
D=\frac{1}{2}\lim_{t \to \infty } \frac{d}{{dt}} \left( <l(t)^{2}>-<l(t)>^2 \right). 
\end{equation}
Then, substituting into this expression Eqs.~\ref{eq_l1} and \ref{eq_l2}, we obtain
\begin{eqnarray}\label{eq_D}
  D =  \frac{d^2 }{2} & \left\{(u  - w_T  - 2A)\sum\limits_{k = 0}^{ + \infty } \{T_{k,k}^0  + T_{k+1,k}^0 + T_{k,k+1}^1  + T_{k,k}^1 \} - (w_D - w_T)(T_{0,0}^0  + T_{0,0}^1) \right.   \nonumber \\
 & \left.  \frac{1}{4}(u + w_T) + \frac{1}{4} (w_D - w_T)(b_{0,0} + b_{0,0}^1) \right \}.  
\end{eqnarray}
Note, that $T_{0,0}^0 = T_{0,0}^1 = T_0$ and for sum of all $T_{k,m}$ we have from Eq.~\ref{T}
\begin{equation}
\sum\limits_{k = 0}^{+ \infty}  \{ T_{k,k}^0  + T_{k+1,k}^0 + T_{k,k+1}^1  + T_{k,k}^1 \}  = \frac{2}{1-q} \left \{ T_0  + \frac{1}{4} \frac{u  +  w_D q }{w_T  + r_h} \right \}.
\end{equation}
Finally, after some algebraic transformations of Eqs.~\ref{eq_A} and \ref{eq_D}, we derive the final expression for the growth velocity, $V$ and dispersion, $D$, which are given in Eqs.~\ref{vel} and \ref{disp}  in Section II. Note that the constant $T_0$ cancels out in the final equation.

The mean  size of ATP-cap can be calculated as
\begin{equation}\label{capn1}
 <N_{cap}> = \sum\limits_{k = 0}^{+ \infty} {2k ( b_{k,k}^0 + b_{k,k}^1 ) + (2k+1) ( b_{k+1,k}^0 + b_{k,k+1}^1 )} = \sum\limits_{k = 0}^{+ \infty} {k q^k (1-q)} = \frac{q}{1-q}.
\end{equation}
The average relative fluctuation in the size of the ATP-cap, by definition, is given 
\begin{equation}\label{capfldf}
\frac{<N_{cap}^{2}> -<N_{cap}>^{2}}{<N_{cap}>^{2}} = \frac{\sigma^2}{<N_{cap}>^{2}},
\end{equation}
where
\begin{eqnarray}\label{capn2}
<N_{cap}^{2}> =  \sum\limits_{k = 0}^{+ \infty} {(2k)^2 ( b_{k,k}^0 + b_{k,k}^1 ) + (2k+1)^2 ( b_{k+1,k}^0 + b_{k,k+1}^1 )} \nonumber \\
=  \sum\limits_{k = 0}^{+ \infty} {k^2 q^k (1-q)} = \frac{q + 4q^2 - 3q^3}{(1-q)^{2}}.
\end{eqnarray}
Then from Eqs.~\ref{capn2} and \ref{capn1} we have,
\begin{equation}\label{capfl}
\frac{\sigma^2}{<N_{cap}>^{2}} = \frac{1}{q} + 3(1-q) \quad \ge 1.
\end{equation}

\clearpage

\section*{Figure Legends}
\begin{figure}
 \begin{center}
   \includegraphics*[width=3.25in]{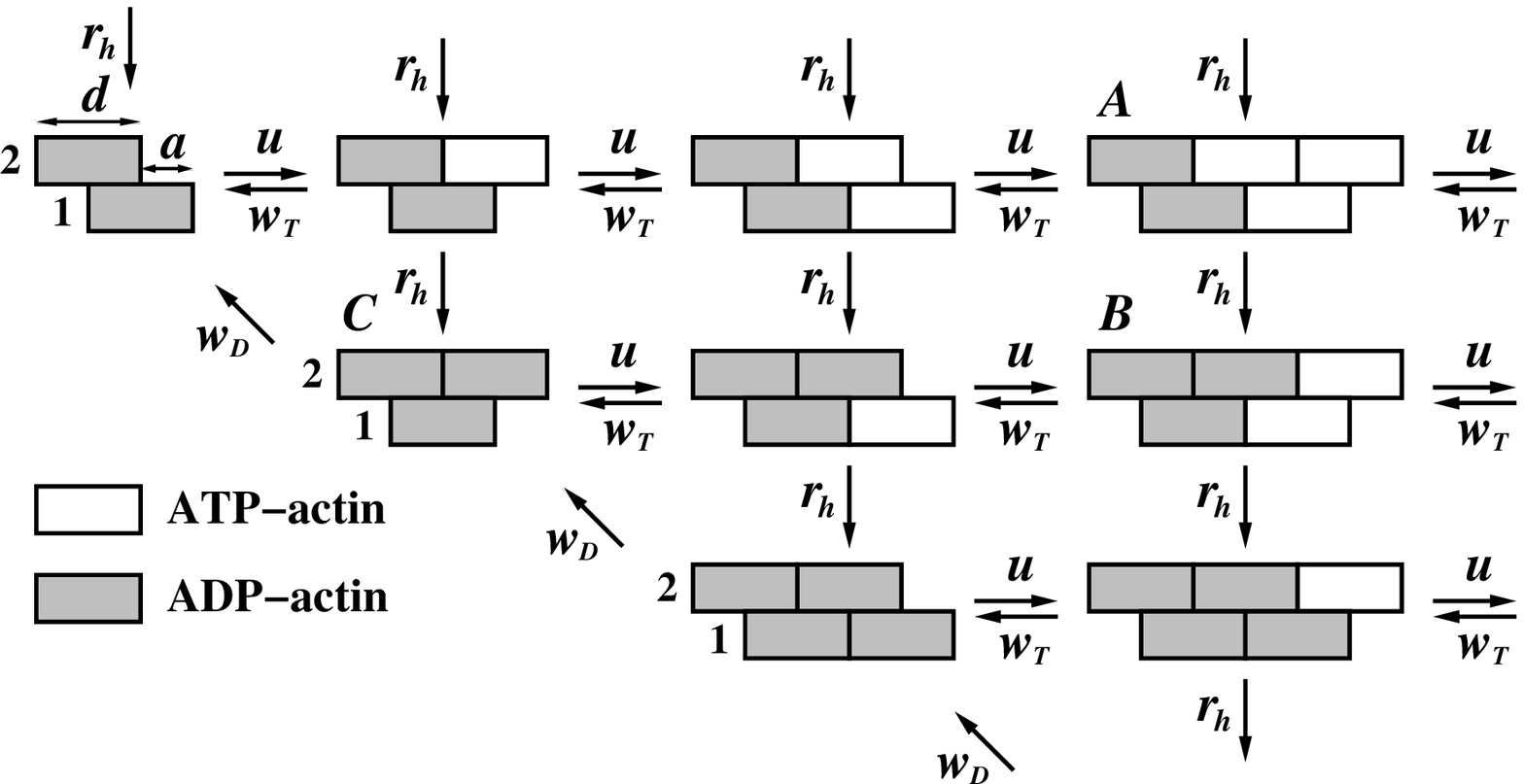}
   \caption{}
   \label{Fig1}
 \end{center}
\end{figure}

\subsubsection*{Figure~\ref{Fig1}.}

Schematic picture of polymer configurations and possible transitions in the vectorial model of the single actin filament's growth. The size of the monomer subunit is $d$, while $a$ is a shift between the parallel protofilaments equal to a half of the monomer size. Two protofilaments are labeled 1 and 2. The transition rates and labels to some of the configurations are explained in the text.

\subsubsection*{Figure~\ref{Fig2}.}

Comparison of the growth velocities for the barbed end of the single  actin filament as a function of free monomeric actin concentration for the random and for the vectorial ATP hydrolysis mechanisms. A vertical dashed line indicates the transition point $c'$ (in the vectorial hydrolysis). It separates  the dynamic regime I (low concentrations) from regime II (high concentrations). The kinetic rate constants used for calculations of the velocities  are taken from the Table 1.

\subsubsection*{Figure~\ref{Fig3}.}

Dispersion of the length of the single actin filament as a function of free  monomer concentration for the barbed end and for the pointed end (the vectorial mechanisms). The kinetic rate constants are taken from the Table 1.  Vertical dashed lines indicate the critical concentrations $c_{crit}$ for the barbed and the pointed ends; thin solid line corresponds to the  concentration of the treadmilling, which is also almost the same as the transition point for the barbed end.  The transition point for the pointed end is at 0.85 $\mu$M. Total dispersion is a sum of the independent contributions for each end of the filament.

\subsubsection*{Figure~\ref{Fig4}.}

The size of ATP cap as a function of free monomer concentration for the barbed end of the single actin filament within the random and the vectorial mechanisms of ATP hydrolysis. Thick solid lines describe the vectorial mechanism, while dotted lines correspond to the random mechanism. The kinetic parameters for constructing curves {\bf b} and {\bf d} are taken from the Table 1. For curves {\bf a} and {\bf c} the kinetic rate constants are the same with the exception of the smaller hydrolysis rate $r_{h}=0.03$ s$^{-1}$.   Vertical dashed line and  thin solid line indicate the transition point $c'$ and the critical concentration $c_{crit}$, respectively, for the curve  {\bf b}.

\subsubsection*{Figure~\ref{Fig5}.}

The fractions of the capped configurations $q$ for the barbed end of the single actin filament as a function of free monomer concentration. The results for two mechanisms of hydrolysis are presented.  Thick solid lines describe the vectorial mechanism, while dotted lines correspond to the random mechanism.  The kinetic parameters for constructing curves {\bf b} and {\bf d} are taken from the Table 1. For curves {\bf a} and {\bf c} the kinetic rate constants are the same with the exception of the smaller hydrolysis rate $r_{h}=0.03$ s$^{-1}$.

\clearpage

\clearpage

\begin{figure}
 \begin{center}
   \includegraphics*[width=3.25in]{acthydr3}
   \caption{}
   \label{Fig2}
 \end{center}
\end{figure}

\clearpage 

\begin{figure}
 \begin{center}
   \includegraphics*[width=3.25in]{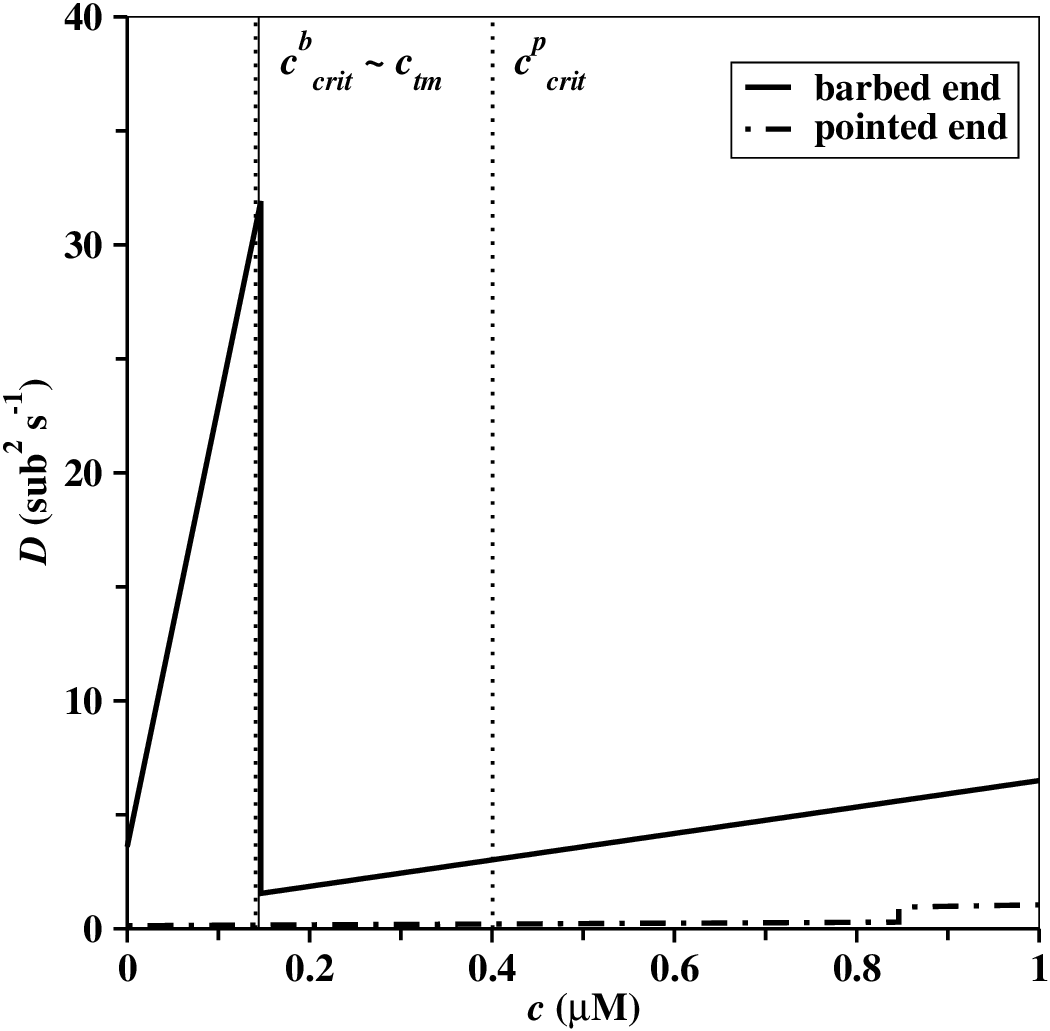}
    \label{Fig3}
    \caption{}
 \end{center}
\end{figure}

\clearpage

\begin{figure}
 \begin{center}
  \includegraphics*[width=3.25in]{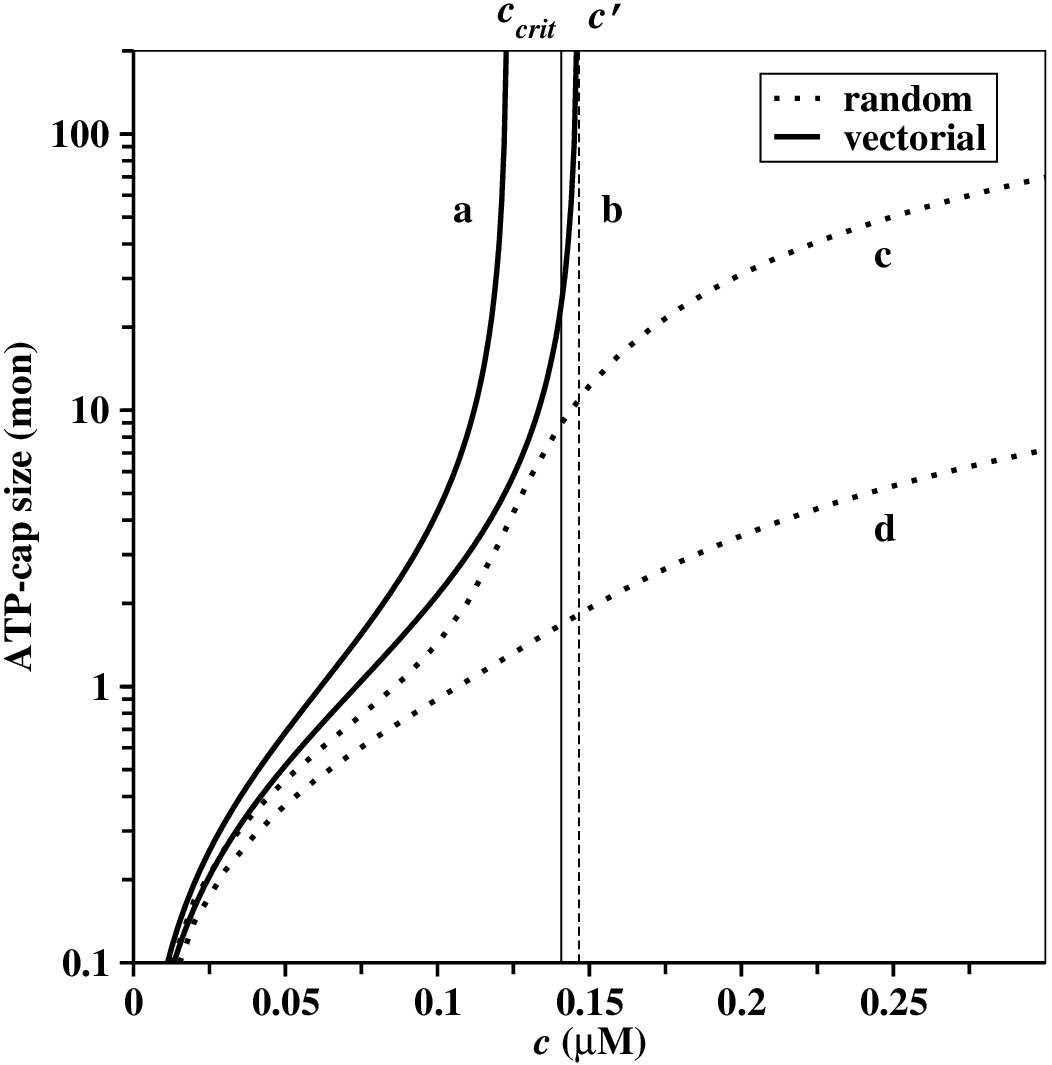}
  \caption{}
  \label{Fig4}
  
 \end{center}

\end{figure}

\clearpage

\begin{figure}
 \begin{center}
  \includegraphics*[width=3.25in]{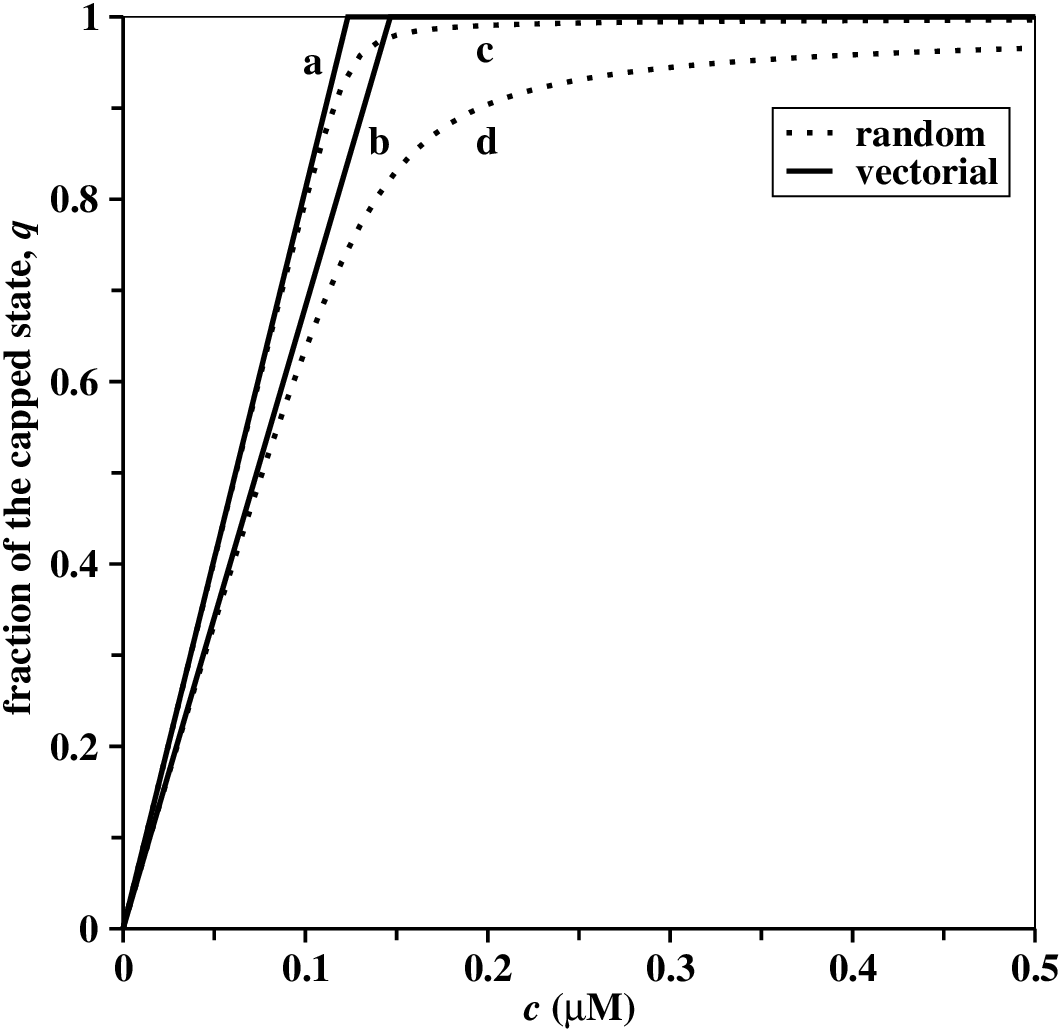}
  \caption{}
  \label{Fig5}
 \end{center}

\end{figure}

\end{document}